\documentclass{amsart}

\usepackage{amssymb}
\usepackage{amsfonts}
\usepackage{amsmath}
\usepackage{lastpage}
\usepackage[utf8]{inputenc}
\usepackage[legalpaper,bookmarks=true,colorlinks=true,linkcolor=blue,citecolor=blue]{hyperref}
\usepackage{graphicx}%
\usepackage{fancyhdr}
\usepackage{color}
\usepackage[mathlines]{lineno}
\usepackage{lscape}
\usepackage{epsfig}
\usepackage{natbib}
\usepackage{booktabs}
\usepackage{geometry}
\usepackage{tgbonum}
\fontfamily{qcr}\selectfont

\newtheorem{theorem}{Theorem}
\theoremstyle{plain}

\numberwithin{equation}{section}

\begin{document}
\Large
\title[The Discrete Zenga Inequality Measure]{A multinomial Asymptotic Representation of Zenga's Discrete Index, its Influence Function and Data-driven Applications}

\begin{abstract}In this paper, we consider the Zenga index, one of the most recent inequality index. We keep the finite-valued original form and address the asymptotic theory. The asymptotic normality is established through a multinomial representation. The Influence function is also given. Th results are simulated and applied to Senegalese data.\\

\bigskip\noindent 
\author{$^{\dag}$ Tchilabalo Atozou Kpanzou}
\author{$^{\dag\dag}$ Diam Ba}
\author{$^{\dag\dag\dag}$ Pape Djiby Mergane}
\author{$^{\dag\dag\dag\dag}$ Gane Samb Lo}

\noindent $^{\dag}$ Tchilabalo Abozou Kpanzou (corresponding author).\\
Kara University, Kara, Togo.\\
Email : kpanzout@gmail.com\\ 

\noindent $^{\dag\dag}$ Diam Ba\\
LERSTAD, Gaston Berger University, Saint-Louis, S\'en\'egal.\\
Email : diamba79@gmail.com.\\

\noindent $^{\dag\dag\dag}$ Pape Djiby Mergane\\
LERSTAD, Gaston Berger University, Saint-Louis, S\'en\'egal.\\
Email : mergane@gmail.com.\\

\noindent $^{\dag\dag\dag\dag}$ Gane Samb Lo.\\
LERSTAD, Gaston Berger University, Saint-Louis, S\'en\'egal (main affiliation).\newline
LSTA, Pierre and Marie Curie University, Paris VI, France.\newline
AUST - African University of Sciences and Technology, Abuja, Nigeria\\
gane-samb.lo@edu.ugb.sn, gslo@aust.edu.ng, ganesamblo@ganesamblo.net\\
Permanent address : 1178 Evanston Dr NW T3P 0J9,Calgary, Alberta, Canada.\\

\noindent \textbf{keywords and phrases}. Inequality measures; Asymptotic behaviour; Asymptotic representations; functional empirical proces.\\

\end{abstract}

\maketitle

\section{Introduction} \label{zenzaDist01}

\noindent Over the years, a number of measures of inequality have been developed. Examples include the generalized entropy, the Atkinson, the Gini, the quintile share ratio and the Zenga measures (see e.g. \cite{Zenga1984} and \cite{Zenga1990}), \cite{CowellandFlachaire2007}; \cite{Cowelletal2009}; \cite{HulligerandSchoch2009}. Recently, \cite{merganelo2013} gathered a significant number of inequality measures under the name of Theil-like family. Such inequality measures are very important in capturing inequality in income distributions. They also have applications in many other branches of Science, e.g. in ecology (see e.g. \cite{Magurran1991}), sociology (see e.g. \cite{Allison1978}), demography (see e.g. \cite{White1986}) and information science (see e.g. \cite{Rousseau1993}).\\

\noindent The inequality measure of \cite{zenga2006} is one of the most recent one. It is receiving a considerable attention from researchers for its novelty indeed, but for its interesting properties. Papers dealing with that measure cover theoretical aspects including asymptotic theory and statistical inference (\cite{Greselinetal2010b}, \cite{emad1999}) and applied works to income data (\cite{Greselinetal2010a}), etc.\\

\noindent In this paper, we focus on the discrete form as introduced by \cite{zenga2006}. We justify the asymptotic study of the discrete and finite form by a number of reasons. In some situations, only aggregated data exists. Although this is hardly conceivable today, it is still possible and it is highly probable that the researcher does not have access to the original data and has in hand only data in form of frequency tables. Some other times, frequency tables may be available while the full data is destroyed or lost. Right now, in Gambia, health data collected from the health centers are stored in 
daily books and the national health direction extract frequency tables from those books and this type of data is the only one available in their computerized system. So one of the main reason to work on the finite discrete data is the lack of accessibility to the full data for one reason or another. The second main is that an asymptotic theory on such king of data will give the structure of the limit results with also no severe conditions. By replacing the discrete finite probability law of the aggregated data by a general probability law, we get the precise general asymptotic case. From that simplified study, we see what might be expected in general theory before we proceed it.\\   

\noindent Here, we suppose that the full data has been summarized into a frequencies table of the form

\begin{table}
	\centering
		\begin{tabular}{ccc}
		\hline\hline
		classes	$(c_{i-1},c_i)$ & Represents  $x_i^{}*$ &  frequencies $n_i$ \\
		\hline \hline
		$(c_{0},c_1)$ &   $x_1^{\ast}$ &  frequencies $n_1$ \\
		$(c_{1},c_2)$ &   $x_2^{\ast}$ &  frequencies $n_2$ \\
		
		$\vdots$ & $\vdots$ &  $$\vdots$$ \\
		
		$(c_{m-1},c_m)$ &   $x_m^{\ast}$ &  frequencies $n_m$ \\
		\hline \hline
		Total  &   $x_i^{\ast}$ &  $n$ \\
		\hline \hline
		\end{tabular}
	\caption{Frequencies Tables}
	\label{tabFrequent}
\end{table}

\bigskip \noindent  Each class $(c_{i-1},c_i)$ in Table \ref{tabFrequent} is represented by a single point $x_i^{\ast}$, usually taken as the middle of class $x_i^{\ast}=(c_{i-1}+c_i)/2$ (other possible choices are the mean of the median of observation falling in the class). So we may adopt approximatively reconstitute the $n\geq 1$ data as follows

$$
\underbrace{x_1^\ast}_{n_1 \ times} \cdots \underbrace{x_j^\ast}_{n_j \ times} \cdots \underbrace{x_m^\ast}_{n_m \ times}
$$ 

\noindent In the sequel, we suppose that the data itself is discrete and takes a pre-determined number of $m$ value. First, we will give an asymptotic theory which will be given in the form of representation in multinomial laws, in opposite to representation in Brownian Bridges in the general case. Next, the influence function will be derived by direct computations and this usually allows to find again the asymptotic variance and some times as in our case, to find a different but equivalent expression of that variance.\\

\noindent The works presented here will be applied to incomes available in an aggregated form. At the same time, they serve as a paving way to a more general approach.\\

\noindent Let us suppose that the income variable $X$\ is discrete and takes the $m$\ ($m>1$) ordered values $0=-\infty <x_{1}<...<x_{m}<x_{m+1}=+\infty $ with the probabilities $p_{j}>0$ , $j\in \{1,...,m\}$ with $p_{1}+p_{2}+...+p_{m}=1$. If the income continuously observed, we have a sequence of random replications $%
X_{1},X_{2},...$ defined on the same probability space $(\Omega, \mathcal{A}, \mathbb{P})$. For each $n\geq 1, $ the empirical distribution of $X$ on the sample is characterized by the
empirical frequencies

\begin{equation*}
n_{0}=0,\ \ n_{j}=\#\{h\in \{1,...,n\},\ \ X_{h}=x_{j}\},\ \ j\in \{1,...,m\},
\end{equation*}

\noindent and their normalized and cumulative forms respectively

\begin{equation*}
f_{0}=0, \ f_{j}=\frac{n_{j}}{n}, \ j\in \{1,...,m\}
\end{equation*}

\noindent and 
\begin{equation*}
n_{0}^{\ast }=f_{0}^{\ast}=0,n_{j}^{\ast}=\sum_{h=1}^{j} n_{h}, \
f_{j}^{\ast}=\sum_{h=1}^{j} f_{h},\ \ j\in \{1,...,m\},
\end{equation*}

\noindent with

\begin{equation*}
\sum_{j=1}^{m}n_{j}=n,\ \ \sum_{j=1}^{m}f_{j}=1,\ \ n_{m}^{\ast }=n,\ \
f_{m}^{\ast }=1.
\end{equation*}

\bigskip \noindent We also define

\begin{equation*}
p_{0}^{\ast }=0, \ \ p_{j}^{\ast}=\sum_{h=1}^{j} p_{h}, \ p_{m}^{\ast}=1.
\end{equation*}

\bigskip \noindent The empirical and discrete \cite{zenga2006}'s index is given by

\begin{equation*}
Z_{d,n}=1-\sum_{j=1}^{m-1}f_{j}\frac{(n_{j}^{\ast }/n)^{-1}\sum_{1\leq h\leq j}n_{h}x_{h}}{(1-(n_{j}^{\ast }/n))^{-1}\sum_{j+1\leq h\leq m}n_{h}x_{h}},
\end{equation*}

\bigskip \noindent which is obtained by summing Formula (3.1) in \cite{zenga2006} over $j\in \{1,...,m\}$ and presented as a synthetic measure of inequality. The empirical cumulative distribution function (\text{cdf}) based on the sample of size $n\geq 1$ is

\begin{equation*}
F_{n}(x)=\frac{1}{n}\sum_{h=1}^{m}n_{h}1_{[x_{h},x_{h+1}[}(x),\ \ x\in 
\mathbb{R}
\end{equation*}

\noindent and is the non-parametric estimator of the true (\text{cdf})

\begin{equation*}
F_{n}(x)=\sum_{h=1}^{m}p_{j}1_{[x_{h},x_{h+1}[}(x), \ \ x\in \mathbb{R}
\end{equation*}

\bigskip \noindent We also have the empirical probability generated by the
sample is given by 
\begin{equation*}
P_{X,n}(A)=\frac{1}{n}\sum_{j=1}^{m}1_{A}(x_{j})
\end{equation*}

\noindent We may express $Z_{n,d}$ in terms of the empirical probability
measure by 

\begin{equation*}
Z_{d,n}=1-\sum_{j=1}^{m-1}\mathbb{P}_{X,n}(x_{j})\frac{\left( \int
1_{]0,x_{j}]}(t)d\mathbb{P}_{X,n}(t)\right) ^{-1}\left( \int
t1_{]0,x_{j}]}(t)d\mathbb{P}_{X,n}(t)\right) }{\left( \int 1_{]x_{j},+\infty
\lbrack }(t)d\mathbb{P}_{X,n}(t)\right) ^{-1}\left( \int t1_{]x_{j},+\infty
\lbrack }(t)d\mathbb{P}_{X,n}(t)\right) }.
\end{equation*}

\bigskip \noindent Finally by considering the discrete measure $\nu =\sum_{1\leq j\leq n}\delta _{x_{j}}$, where $\delta _{x_{j}}$ is the Dirac measure concentrated at $x_{j}$\ with mass one, we may also write

\begin{equation*}
Z_{d,n}=1-\int \frac{\left( \int 1_{]0,s]}(t)d\mathbb{P}_{X,n}(t)\right)
^{-1}\left( \int t1_{]0,s]}(t)d\mathbb{P}_{X,n}(t)\right) }{\left( \int
1_{]s,+\infty \lbrack }(t)d\mathbb{P}_{X,n}(t)\right) ^{-1}\left( \int
t1_{]s,+\infty \lbrack }(t)d\mathbb{P}_{X,n}(t)\right) }\mathbb{P}_{X,n}(s)d\nu (s).
\end{equation*}

\bigskip \noindent It is clear, by the convergence in law of the sequence of probability measures $\mathbb{P}_{X,n}$ to the $\mathbb{P}_{X}=\mathbb{P}X^{-1}$ (the probability law of $X$), we see that $Z_{n,d}$ converges to

\begin{equation*}
Z_{d}=1-\int \frac{\left( \int 1_{]0,x_{j}]}(t)d\mathbb{P}_{X}(t)\right)
^{-1}\left( \int t1_{]0,x_{j}]}(t)d\mathbb{P}_{X}(t)\right) }{\left( \int
1_{]x_{j},+\infty \lbrack }(t)d\mathbb{P}_{X}(t)\right) ^{-1}\left( \int
t1_{]x_{j},+\infty \lbrack }(t)d\mathbb{P}_{X}(t)\right) }\mathbb{P}%
_{X}(s)d\nu (s).
\end{equation*}

\bigskip \noindent In this simple setting, the convergence are easily justified because of the finiteness of the summations and of the functions. In terms of \text{cdf} and on mathematical expectation, we have

\begin{equation*}
Z_{d}=1-\int_{x_{1}}^{x_{m}}\frac{\frac{1}{F(s)}\int_{0}^{s}td\mathbb{P}_{X}(t)}{\frac{1}{1-F(s)}\int_{s}^{\infty }td\mathbb{P}_{X}(t)}\mathbb{P}_{X}(s)d\nu (s).(X)
\end{equation*}

\bigskip \noindent The integral in the last expression should be read as 

\begin{equation*}
\int_{x_{1}}^{x_{m}-}\frac{\frac{1}{F(s)}\int_{0}^{s}td\mathbb{P}_{X}(t)}{%
\frac{1}{1-F(s)}\int_{s}^{\infty }td\mathbb{P}_{X}(t)}\mathbb{P}_{X}(s)d\nu
(s)=\int \ 1_{[x_{1},x_{m}[}(s)\frac{\frac{1}{F(s)}\int_{0}^{s}td\mathbb{P}%
_{X}(t)}{\frac{1}{1-F(s)}\int_{s}^{\infty }td\mathbb{P}_{X}(t)}\mathbb{P}%
_{X}(s)d\nu (s),
\end{equation*}

\bigskip \noindent so that neither $1-F(s)$\ nor $F(s)$ never vanishes on the integration domain.\\

\noindent On one side, we are going to draw an asymptotic normality theory of $Z_{n,d}$ using the $m$-multivariate binomial laws. On an other side, the sensitivity of a statistic $T(F)$ and the impact of extreme observations on it are also two recurrent questions in the research in the field (see \cite{CowellandFlachaire2007})\newline

\noindent In that context, the asymptotic variance of the plug-in estimator $T(F_{n})$ of statistic $T(F)$ is of the form $\sigma ^{2}=\int L(x,T(F))^{2}dF(x)$. From this, we may say that the influence function behaves in nonparametric estimation as the score function does in the parametric setting (See \cite{wasserman2006}, page 19). To define the notion of \textit{IF}, 
Let us consider the contaminated probability law $\mathbb{P}_{X}^{-(\varepsilon )}$ of $\mathbb{P}_{X}$ at $x$ with mass $\varepsilon>0$ by

\begin{equation}
\mathbb{P}_{X}^{(\varepsilon )}=(1-\varepsilon )\mathbb{P}_{X}+\varepsilon \delta _{x}. \label{probContam}
\end{equation}

\bigskip \noindent and a functional of $\mathbb{P}_{X}$, namely $T(\mathbb{P}_{X})$. The influence function of the functional $T$ at $x$, if it exists, is given by 
\begin{equation}
IF(T,x)=\lim_{\varepsilon \rightarrow 0}\frac{T(\mathbb{P}_{X}^{(\varepsilon )})-T(\mathbb{P}_{X})}{\varepsilon}. \ \label{defIF}
\end{equation}

\bigskip \noindent The previous remarks motivate us to derive the \textit{IF} function of $Z_d(\mathbb{P}_X)$ and to compare it with the asymptotic variance the plug-in Zenga's estimator.\\

\noindent Before we proceed to our a task, we point out that asymptotic normality results for Zenga's index are available in the literature, among them those of \cite{Greselinetal2010b} and \cite{emad1999}. We will come back to these results in the coming paper where we deal with other version of asymptotic versions in the general case.\\

\noindent Here is how is organized the paper, we give our asymptotic results as described above in Section \ref{zengaDist02} in Theorems \ref{theoD1} and 
\ref{theoD2}. Section \ref{zengaDist03} is devoted to simulation studies and data-driven application to Senegalese Data. A conclusion and perspectives section ends the paper.\\

\section{Asymptotic Theory for the discrete Zenga measure} \label{zengaDist02}

\bigskip \noindent \textbf{(A) - Asymptotic normality}.\\

\bigskip Let begin by the following reminder. For each $n\geq 1$, the random vector $(n_1, ....,n_m)$ follows a $m$-dimensional multimonial law of
parameters $n\geq 1$ and $p=(p_1,...,p_m)^t$. In such a case a classical result of weak convergence (See \cite{ips-wcia-ang}, for example,, as $n\rightarrow +\infty$, is the following

\begin{eqnarray*}
\left(\frac{n_1-np_1}{\sqrt{np_1}},\cdots,\frac{n_m-np_m}{\sqrt{np_m}}%
\right)^t&\equiv& \left(N_{1,n},\cdots,N_{m,n}\right)^t \\
&\rightsquigarrow& Z=(Z_1,\cdots,Z_m)^t \sim \mathcal{N}_m(0,\Sigma),
\end{eqnarray*}

\bigskip \noindent the variance-covariance matrix $\Sigma=(\sigma_{h,k})_{1\leq
k,k\leq m}$ of $Z$ is defined, for $(h,k) \in \{1,...,m\}^2$, $h\neq k$, by

\begin{equation*}
\sigma_{hh}=\mathbb{E}(Z_h^2)=1-p_h \ and \ \sigma_{hk}=\mathbb{E}(Z_hZ_k)=-\sqrt{p_hp_k}. 
\end{equation*}

\bigskip \noindent We invoke the Skorohod-Wichura Theorem (See \cite{wichura1970}) to suppose that $Z$ is defined on the same probability space and that

\begin{equation*}
\left(N_{1,n},\cdots,N_{m,n}\right)^t \rightarrow_{\mathbb{P}} Z, \ as \
n\rightarrow +\infty.
\end{equation*}

\bigskip \noindent Let us give some notation. \noindent Define vectors $C=(c_{1},...,c_{m})^{t}$ such that

\begin{equation*}
c_{j}=\sqrt{p_{j}}\frac{(1/p_{j}^{\ast })\mu _{(j)}}{(1/(1-p_{j}^{\ast
}))\mu ^{(j)}}1_{\left( j\neq m\right)} ,j\in \{1,...,m\},
\end{equation*}

\bigskip \noindent for $j\in \{1,...m-1\}$, $i\in \{1,2\}$, $%
D_{j,i}=(d_{j,i,1},...,d_{j,i,m})^{t}$ such that

\begin{equation*}
d_{j,1,h}=\left( x_{h}\sqrt{p_{h}}\right) 1_{(h\leq j)}, \ \
d_{j,2,h}=-\left( x_{h}\sqrt{p_{h}}\right) 1_{(h\geq j+1)}
\end{equation*}

\begin{equation*}
\gamma _{j,1}=p_{j}\frac{(1/p_{j}^{\ast })}{(1/(1-p_{j}^{\ast }))\mu ^{(j)}}, \ \ \gamma _{j,2}=p_{j}\frac{(1/p_{j}^{\ast })}{(1/(1-p_{j}^{\ast})}\frac{\mu_{(j)}}{\left(\mu^{(j)}\right)^2}
\end{equation*}

\bigskip \noindent and let $E_{j}=(e_{j,1},...,e_{j,m})^{t}$ be the vector defined
by its components as follows

\begin{equation*}
e_{j,h}=- \left(\sqrt{p_{h}}\right) 1_{(h\leq j)}. 
\end{equation*}

\noindent Finally, let us defined

\begin{equation*}
-H=C+\sum_{j=1}^{m-1}\left(\gamma_{j,1}D_{j,1}+\gamma_{j,2}D_{j,2}+\left(p_j^{\ast}\right)^{-2}E_j\right).
\end{equation*}

\begin{theorem} \label{theoD1} Under the notation given above, we have, as $n\rightarrow +\infty$,

\begin{equation*}
\sqrt{n}(Z_{d,n}-Z_{d}) \rightsquigarrow \mathcal{N}_m\left(0, H^t\Sigma H\right). \ \Diamond 
\end{equation*}
\end{theorem}

\bigskip \noindent \textbf{Proof of Theorem \ref{theoD1}}. Let us fix \ $n\geq 1$. We have

\begin{equation*}
Z_{n,d}=1-\sum_{j=1}^{m-1}\frac{n_{j}}{n}\left( \frac{n}{n_{j}^{\ast }}-1\right) \frac{\sum_{1\leq h\leq j}n_{h}x_{h}}{\sum_{j+1\leq h\leq
m}n_{h}x_{h}}.
\end{equation*}

\noindent We define
\begin{equation*}
Z_{d,n}^{\ast }=\sum_{j=1}^{m-1}\frac{n_{j}}{n}\left( \frac{n}{n_{j}^{\ast }}%
-1\right) \frac{\sum_{1\leq h\leq j}n_{h}x_{h}}{\sum_{j+1\leq h\leq
m}n_{h}x_{h}}.
\end{equation*}

\noindent and for $1\leq j\leq m-1,$%
\begin{equation*}
\mu_{(j)}=\sum_{h=1}^{j}p_{h}x_{h}\ \ and \ \mu
^{(j)}=\sum_{h=j+1}^{m}p_{h}x_{h}.
\end{equation*}

\noindent We have
\begin{eqnarray*}
&&\frac{\sum_{1\leq h\leq j}n_{h}x_{h}}{\sum_{j+1\leq h\leq m}n_{h}x_{h}}-%
\frac{\mu _{(j)}}{\mu ^{(j)}} \\
&=&\frac{\sum_{1\leq h\leq j}n_{h}x_{h}}{\sum_{j+1\leq h\leq m}n_{h}x_{h}} -%
\frac{n\mu _{(j)}}{\sum_{j+1\leq h\leq m}n_{h}x_{h}} \\
&+&\frac{n\mu _{(j)}}{\sum_{j+1\leq h\leq m}n_{h}x_{h}}-\frac{\mu_{(j)}}{\mu
^{(j)}} \\
&=&\frac{\sum_{h=1}^{j}x_{h}N_{h,n}\sqrt{p_{h}}}{\sqrt{n}\sum_{j+1\leq h\leq
m}n_{h}x_{h}/n}-\frac{\mu_{(j)}\sum_{h=j+1}^{m}x_{h}N_{h,n}\sqrt{p_{h}}}{%
\sqrt{n}\mu ^{(j)}\left( \sum_{j+1\leq h\leq m}n_{h}x_{h}/n\right) }.
\end{eqnarray*}

\noindent Then
\begin{eqnarray*}
&&Z_{d,n}^{\ast} \\
&=&\sum_{j=1}^{m-1}\frac{n_{j}}{n}\left( \frac{n}{n_{j}^{\ast }}-1\right) 
\frac{\mu_{(j)}}{\mu^{(j)}} \\
&+&\frac{1}{\sqrt{n}}\sum_{j=1}^{m-1}\frac{n_{j}}{n}\left( \frac{n}{%
n_{j}^{\ast }}-1\right) \left( \frac{\sum_{h=1}^{j}x_{h}N_{h,n}\sqrt{p_{h}}}{%
\sum_{j+1\leq h\leq m}n_{h}x_{h}/n}-\frac{\mu_{(j)}%
\sum_{h=j+1}^{m}x_{h}N_{h,n}\sqrt{p_{h}}}{\mu ^{(j)}\left(\sum_{j+1\leq
h\leq m}n_{h}x_{h}/n\right) }\right) \\
&=&:Z_{d,n}^{\ast}(1)+R_{n}(1,1)
\end{eqnarray*}

\bigskip \noindent We also have

\begin{eqnarray*}
\left( \frac{n}{n_{j}^{\ast }}-1\right) -\left( \frac{1}{p_{j}^{\ast }}%
-1\right) &=&\left( \frac{n}{n_{j}^{\ast }}-1\right) -\left( \frac{n}{%
\sum_{h=1}^{j}np_{h}}-1\right) \\
&=&-\frac{\sum_{h=1}^{j}n_{h}-\sum_{h=1}^{j}p_{h}}{\left(\sum_{h=1}^{j}p_{h}%
\right)\left(\sum_{h=1}^{j}n_{h}\right)} \\
&=&-\frac{1}{\sqrt{n}}\frac{\sum_{h=1}^{j}\sqrt{p_{h}}N_{h,n}}{%
\left(\sum_{h=1}^{j}p_{h}\right)\left(\sum_{h=1}^{j}n_{h}/n\right)}.
\end{eqnarray*}

\noindent This leads to

\begin{eqnarray*}
Z_{d,n}^{\ast}(1) &=&\sum_{j=1}^{m-1}\frac{n_{j}}{n}\left( \frac{1}{%
p_{j}^{\ast }}-1\right) \frac{\mu _{(j)}}{\mu ^{(j)}}-\sum_{j=1}^{m-1}\frac{n_{j}}{n}\frac{n\sqrt{n}\sum_{h=1}^{j}\sqrt{p_{h}}N_{h,n}}{\left(
\sum_{h=1}^{j}n_{h}\right) \left( \sum_{h=1}^{j}np_{h}\right) }\frac{\mu
_{(j)}}{\mu ^{(j)}} \\
&=&:Z_{d,n}^{\ast}(2)+R_{n}(1,2)
\end{eqnarray*}

\bigskip \noindent Finally, we have

\begin{eqnarray*}
Z_{d,n}^{\ast }(2) &=&\sum_{j=1}^{m-1}p_{j}\left( \frac{1}{p_{j}^{\ast }}%
-1\right) \frac{\mu _{(j)}}{\mu ^{(j)}}+\frac{1}{n}\sum_{j=1}^{m-1}\frac{%
\sqrt{np_{j}}N_{j,n}}{{}}\left( \frac{1}{p_{j}^{\ast }}-1\right) \frac{\mu
_{(j)}}{\mu ^{(j)}} \\
&=&\sum_{j=1}^{m-1}p_{j}\left( \frac{1}{p_{j}^{\ast }}-1\right) \frac{\mu
_{(j)}}{\mu ^{(j)}}+\frac{1}{\sqrt{n}}\sum_{j=1}^{m-1}\sqrt{p_{j}}%
N_{j,n}\left( \frac{1}{p_{j}^{\ast }}-1\right) \frac{\mu _{(j)}}{\mu ^{(j)}}%
\text{ \ \ (L2)} \\
&=&\sum_{j=1}^{m-1}\frac{(1/p_{j}^{\ast })\mu _{(j)}}{(1/(1-p_{j}^{\ast
}))\mu ^{(j)}}+\frac{1}{\sqrt{n}}\sum_{j=1}^{m-1}\sqrt{p_{j}}N_{j,n}\left( 
\frac{1}{p_{j}^{\ast }}-1\right) \frac{\mu _{(j)}}{\mu ^{(j)}} \\
&=&:Z_{d}^{\ast }+R_{n}(3).
\end{eqnarray*}

\bigskip \noindent It is clear that

\begin{equation*}
Z_{d}=1-Z_{d}^{\ast }.
\end{equation*}

\bigskip \noindent We finally get

\begin{equation*}
\sqrt{n}(Z_{d,n}^{\ast }-Z_{d}^{\ast })=\sqrt{n}R_{n}(1)+\sqrt{n}R_{n}(2)+%
\sqrt{n}R_{n}(3).
\end{equation*}

\bigskip \noindent By using the convergence (strong and weak) on binomial
probabilities, we get

\begin{eqnarray*}
&&\sqrt{n}R_{n}(1,1) \\
&=& \sum_{j=1}^{m-1}\frac{n_{j}}{n}\left( \frac{n}{n_{j}^{\ast }}-1\right)
\left(\frac{\sum_{h=1}^{j}\left(x_{h}\sqrt{p_{h}}\right)N_{h,n}}{%
\sum_{j+1\leq h\leq m}n_{h}x_{h}/n} -\frac{\mu_{(j)}\sum_{h=j+1}^{m}%
\left(x_{h}\sqrt{p_{h}}\right)N_{h,n}}{\mu ^{(j)}\left( \sum_{j+1\leq h\leq
m}n_{h}x_{h}/n\right) }\right) \\
&\rightarrow_{\mathbb{P}}& \sum_{j=1}^{m-1}p_{j}\frac{(1/p_{j}^{\ast})}{%
(1/(1-p_{j}^{\ast }))}\left(\frac{\sum_{h=1}^{j}\left(x_{h}\sqrt{p_{h}}%
\right)Z_{h}}{\mu^{(j)}} -\frac{\mu_{(j)}\sum_{h=j+1}^{m}\left(x_{h}\sqrt{%
p_{h}}\right)Z_h}{\left(\mu ^{(j)}\right)^2}\right), \ \ (A1)
\end{eqnarray*}

\bigskip \noindent Next 
\begin{eqnarray*}
\sqrt{n}R_{n}(1,2)&=&-\frac{\sum_{h=1}^{j}\sqrt{p_{h}}N_{h,n}}{%
\left(\sum_{h=1}^{j}p_{h}\right)\left(\sum_{h=1}^{j}n_{h}/n\right)} \\
&\rightarrow_{\mathbb{P}}& -\frac{\sum_{h=1}^{j}\sqrt{p_{h}}Z_h}{%
(p_j^{\ast})^2}. \ \ (A2)
\end{eqnarray*}

\bigskip \noindent and finally

\begin{eqnarray*}
\sqrt{n}R_{n}(3)&=& \sum_{j=1}^{m-1}\sqrt{p_{j}}\left(\frac{1}{p_{j}^{\ast }}%
-1\right) \frac{\mu _{(j)}}{\mu ^{(j)}}N_{j,n} \\
&\rightarrow_{\mathbb{P}}&\sum_{j=1}^{m-1}\sqrt{p_{j}}\frac{(1/p_{j}^{\ast
})\mu _{(j)}}{(1/(1-p_{j}^{\ast }))\mu ^{(j)}}Z_{j}. \ \ (A3)
\end{eqnarray*}

\bigskip \noindent By combining Developments (A1), (A2) and (A3), we get

\begin{eqnarray*}
&&\sqrt{n}(Z_{d,n}^{\ast }-Z_{d}^{\ast }) \\
&\rightarrow & \sum_{j=1}^{m-1}p_{j}\frac{(1/p_{j}^{\ast})}{%
(1/(1-p_{j}^{\ast }))}\left(\frac{\sum_{h=1}^{j}\left(x_{h}\sqrt{p_{h}}%
\right)Z_{h}}{\mu^{(j)}} -\frac{\mu_{(j)}\sum_{h=j+1}^{m}\left(x_{h}\sqrt{%
p_{h}}\right)Z_h}{\left(\mu ^{(j)}\right)^2}\right) \\
&-&\frac{\sum_{h=1}^{j}\sqrt{p_{h}}Z_h}{(p_j^{\ast})^2} \\
&+&\sum_{j=1}^{m-1}\sqrt{p_{j}}\frac{(1/p_{j}^{\ast })\mu _{(j)}}{%
(1/(1-p_{j}^{\ast }))\mu ^{(j)}}Z_{j} \\
&=&\left(\sum_{j=1}^{m-1}\langle \gamma_{j,1}D_{j,1},Z\rangle +\langle
\gamma_{j,2}D_{j,2},Z\rangle + \langle (p_j^{\ast})^{-2} E_{j},Z\rangle\right)
+\langle C,Z\rangle.
\end{eqnarray*}

\bigskip \noindent We conclude that 
\begin{equation*}
\sqrt{n}(Z_{d,n}^{\ast }-Z_{d}^{\ast })\rightarrow _{P}H^{t}Z. \ \square
\end{equation*}

\newpage
\noindent \textbf{(B) - Influence Function of $Z_d$}.\\

\begin{theorem} \label{theoD2} Under the notations given below, the Influence function of $Z_d$ is given, for $x_1\leq x \leq x_m$, by

\begin{eqnarray*}
IF(Z_{d},x) &=&\int \mathbb{P}_{X}(s) \left( \frac{R_{1}(s)}{R_{2}(s)^{2}(1-F(s)})1_{]s,+\infty ]}(x)-\frac{1}{R_{2}(s)F(s)}1_{]0,s]}(x)\right)xd\nu  \\\
&+& \int \mathbb{P}_{X}(s) \left( \frac{R_{1}(s)}{R_{2}(s)F(s)}1_{]0,s]}(x)-\frac{R_{1}(s)}{R_{2}(s)(1-F(s))}1_{]s,+\infty ]}(x)\right) d\nu  \\
&-&\int \delta _{x}(s)\frac{R_{1}(s)}{R_{2}(s)}d\nu +\int \mathbb{P}_{X}(s)%
\frac{R_{1}(s)}{R_{2}(s)}d\nu .
\end{eqnarray*}
\end{theorem}

\bigskip \noindent \textbf{Proof of Theorem \ref{theoD2}}. Let us write, for $s \in \mathcal{R}$,

\begin{equation*}
R_{1}(s)=R_{1}(s,\mathbb{P}_{X})=\frac{\int t1_{]0,s]}(t)d\mathbb{P}_{X}(t)}{\int 1_{]0,s]}(t)d\mathbb{P}_{X}(t)},
\end{equation*}

\noindent and

\begin{equation*}
R_{2}(s)=R_{2}(s,\mathbb{P}_{X})=\frac{\int t1_{]s,+\infty \lbrack }(t)d\mathbb{P}_{X}(t)}{\int ]s,+\infty \lbrack d\mathbb{P}_{X}(t)}.
\end{equation*}

\bigskip \noindent We have

\begin{equation*}
Z_{d}(\mathbb{P}_{X})=Z_{d}=1-\int \frac{R_{1}(s)}{R_{2}(s)}\mathbb{P}_{X}(s)d\nu (s).
\end{equation*}

\bigskip \noindent By using Formula (\ref{probContam}), we have

\begin{equation*}
\frac{d(\mathbb{P}_{X}^{(\varepsilon )}-\mathbb{P}_{X})}{\varepsilon }=-d\mathbb{P}_{X}+d\delta _{x}
\end{equation*}

\bigskip \noindent For short, we write

\begin{equation*}
R_{i}(s,\mathbb{P}_{X})=R_{i}(s)\text{ and }R_{i}(s,\mathbb{P}_{X}^{(\varepsilon )})=R_{i}(s,\varepsilon ),i\in \{1,2\}.
\end{equation*}

\bigskip \noindent We have

\begin{eqnarray*}
Z_{d}(\mathbb{P}_{X}^{(\varepsilon )})-Z_{d}(\mathbb{P}_{X}) &=&-(1-\varepsilon)\int \mathbb{P}_{X}(s)\frac{R_{1}(s,\varepsilon )}{R_{2}(s,\varepsilon )}d\nu -\varepsilon \int \delta _{x}(s)\frac{R_{1}(s,\varepsilon )}{R_{2}(s,\varepsilon )}d\nu\\
&+&\int \mathbb{P}_{X}(s)\frac{R_{1}(s)}{R_{2}(s)}d\nu  \\
&=&-\int \mathbb{P}_{X}(s)\left( \frac{R_{1}(s,\varepsilon )}{R_{2}(s,\varepsilon )}-\frac{R_{1}(s)}{R_{2}(s)}\right) d\nu \\
&+&\varepsilon \int \mathbb{P}_{X}(s)\frac{R_{1}(s,\varepsilon )}{R_{2}(s,\varepsilon )}d\nu -\varepsilon \int \delta _{x}(s)\frac{R_{1}(s,\varepsilon )}{R_{2}(s,\varepsilon )}d\nu .
\end{eqnarray*}


\bigskip \noindent Le us apply the definition of the \textit{IF} as in Formula \eqref{defIF}. Since $\mathbb{P}_{X}^{(\varepsilon )}\rightarrow \mathbb{P}_{X}$ as $\varepsilon \rightarrow 0$ (The convergence being meant as a convergence in law), we have no problem to see that

\begin{eqnarray}\label{use_eq_for_proo}
\lim_{\varepsilon \rightarrow 0}\frac{Z_{d}(\mathbb{P}_{X}^{(\varepsilon )})-Z_{d}(\mathbb{P}_{X})}{\varepsilon } &=& \int \mathbb{P}_{X}(s)\frac{R_{1}(s)}{R_{2}(s)}d\nu-\int \delta _{x}(s)\frac{R_{1}(s)}{R_{2}(s)}d\nu \notag \\
&-&\int \mathbb{P}_{X}(s)\lim_{\varepsilon \rightarrow 0}\frac{1}{\varepsilon }\left( \frac{R_{1}(s,\varepsilon )}{R_{2}(s,\varepsilon )}-%
\frac{R_{1}(s)}{R_{2}(s)}\right) d\nu .
\end{eqnarray}

\bigskip \noindent So we have to find the influence function of $R_{1}(s)/R_{2}(s)$. By formally representing the differentiation of a functional $T(\mathbb{P}_{X})$ by

\begin{equation*}
\frac{\partial T(\mathbb{P}_{X})}{\partial \lambda }
\end{equation*}

\noindent we have that the  influence function of $R_{1}(s)/R_{2}(s)$ is given by

\begin{equation*}
IF(R_{1}(s)/R_{2}(s),x)=\frac{R_{2}(s)\frac{\partial R_{1}(s)}{\partial
\lambda }-R_{1}(s)\frac{\partial R_{2}(s)}{\partial \lambda }}{R_{2}(s)^{2}}.
\end{equation*}

\bigskip \noindent But

\begin{eqnarray*}
R_{1}(s,\varepsilon )-R_{1}(s) &=&\frac{\int t1_{]0,s]}(t)d\mathbb{P}_{X}(t)}{\int 1_{]0,s]}(t)d\mathbb{P}_{X}^{(\varepsilon)}(t)}\\
&-&  \frac{\varepsilon\int t1_{]0,s]}(t)d\mathbb{P}_{X}(t)}{\int 1_{]0,s]}(t)d\mathbb{P}_{X}^{(\varepsilon )}(t)} \\
&+&\frac{\varepsilon \int t1_{]0,s]}(t)d\delta_{x}(t)}{\int 1_{]0,s]}(t)d\mathbb{P}_{X}^{(\varepsilon )}(t)}-\frac{\int t1_{]0,s]}(t)d\mathbb{P}_{X}(t)}{\int 1_{]0,s]}(t)d\mathbb{P}_{X}(t)} \\
&=&\frac{\int t1_{]0,x_{j}]}(t)d(\mathbb{P}_{X}^{(\varepsilon )}(t)-\mathbb{P}_{X}(t))}{\int 1_{]0,s]}(t)d\mathbb{P}_{X}^{(\varepsilon )}(t)} \\
&-&\frac{\int 1_{]0,s]}(t)d(\mathbb{P}_{X}^{(\varepsilon )}(t)-\mathbb{P}_{X}(t))}{\left( \int 1_{]0,s]}(t)d\mathbb{P}_{X}^{(\varepsilon)}(t)\right) \left( \int 1_{]0,s]}(t)d\mathbb{P}_{X}(t)\right) }\int
t1_{]0, s]}(t)d\mathbb{P}_{X}(t).
\end{eqnarray*}

\bigskip \noindent We get

\begin{eqnarray*}
\lim_{\varepsilon \rightarrow 0}\frac{R_{1}(s,\varepsilon )-R_{1}(s)}{\varepsilon } &=&\frac{\int t1_{]0,x_{j}]}(t)d(-\mathbb{P}_{X}(t)+\delta _{x})}{\int 1_{]0,s]}(t)d\mathbb{P}_{X}(t)}\\
&-&\frac{\int 1_{]0,s]}(t)d(-\mathbb{P}_{X}(t)+\delta _{x})}{\left( \int 1_{]0,s]}(t)d\mathbb{P}_{X}(t)\right) ^{2}}\int t1_{]0,s]}(t)d\mathbb{P}_{X}(t) \\
&=&\frac{-\left( \int t1_{]0,s]}(t)d\mathbb{P}_{X}(t)\right)+x1_{]0,s]}(x)}{\int 1_{]0,s]}(t)d\mathbb{P}_{X}(t)}\\
&-&\frac{-\left( \int1_{]0,s]}(t)d\mathbb{P}_{X}(t)\right) +1_{]0,s]}(x)}{\left( \int1_{]0,s]}(t)d\mathbb{P}_{X}(t)\right) ^{2}}\int t1_{]0,s]}(t)d\mathbb{P}_{X}(t)
\end{eqnarray*}

\bigskip \noindent We get
\begin{equation*}
\frac{\partial R_{1}(s)}{\partial \lambda }=-R_{1}(s)+\frac{x1_{]0,s]}(x)}{F(s)}+R_{1}(s)-\frac{R_{1}(s)}{F(s)}1_{]0,s]}(x).
\end{equation*}

\bigskip \noindent By treating $R_{2}(s)$ in the same manner we have (We should not forget that we differentiate in the probability)

\begin{eqnarray*}
\frac{\partial R_{1}(s)}{\partial \lambda } &=&\frac{x1_{]0,s]}(x)}{F(s)}-\frac{R_{1}(s)}{F(s)}1_{]0,s]}(x) \\
\frac{\partial R_{2}(s)}{\partial \lambda } &=&\frac{x1_{]s,+\infty ]}(x)}{1-F(s)}-\frac{R_{2}(s)}{1-F(s)}1_{]s,+\infty ]}(x)
\end{eqnarray*}

\noindent Thus

\begin{eqnarray*}
&&\lim_{\varepsilon \rightarrow 0}\frac{R_{1}(s,\varepsilon )-R_{1}(s)}{\varepsilon }=\left( \frac{1_{]0,s]}(x)}{R_{2}(s)F(s)}-\frac{R_{1}(s)1_{]s,+\infty ]}(x)}{R_{2}^{2}(s)(1-F(s))}\right)x\\
&+&\left( \frac{R_{1}(s)}{R_{2}(s)(1-F(s))}1_{]s,+\infty ]}(x) - \frac{R_{1}(s)}{R_{2}(s)F(s)}1_{]0,s]}(x)  \right);
\end{eqnarray*}

\noindent By replacing this limit with its expression in the equation \eqref{use_eq_for_proo} we get.

\begin{eqnarray*}
\lim_{\varepsilon \rightarrow 0}\frac{Z_{d}(\mathbb{P}_{X}^{(\varepsilon )})-Z_{d}(\mathbb{P}_{X})}{\varepsilon } &=& \int \mathbb{P}_{X}(s)\frac{R_{1}(s)}{R_{2}(s)}d\nu-\int \delta _{x}(s)\frac{R_{1}(s)}{R_{2}(s)}d\nu\\
&+&\int \mathbb{P}_{X}(s) \left( \frac{R_{1}(s)}{R_{2}(s)^{2}(1-F(s)})1_{]s,+\infty ]}(x)-\frac{1}{R_{2}(s)F(s)}1_{]0,s]}(x)\right)xd\nu  \\\
	&+& \int \mathbb{P}_{X}(s) \left( \frac{R_{1}(s)}{R_{2}(s)F(s)}1_{]0,s]}(x)-\frac{R_{1}(s)}{R_{2}(s)(1-F(s))}1_{]s,+\infty ]}(x)\right) d\nu.
\end{eqnarray*}

\bigskip \noindent From this, the proof is directed concluded. $\blacksquare$\\

\section{Data-driven Applications} \label{zengaDist03}

\bigskip \noindent \textbf{Simulation Study}. \\

\noindent \textbf{Quality of the convergence}. We choose a Probability distribution of yearly income supported by $m=10$ points with lower endpoint $x_1=4.515.000$ XOF (9.030 nearly)  and upper endpoint 
$x_m=9.000.000$ XOF(170.490 nearly) , characterized as in Table \ref{tab1}.

\begin{table}[htbp]
	\centering
		\begin{tabular}{lllllll}
		\hline \hline
			values & $x_1$ & $x_2$ & $x_3$ & $x_4$ & $x_5$ & ...\\ 
		\hline \hline
			& 4.515.000 & 13.485.000 & 22.455.000 & 31.425.000 & 40.395.000 &... \\
			$\mathbb{P}(X=x_i)$ & 0.05 & 0.05 & 0.05 & 0.05 &  0.1 & ...\\
		\hline \hline
		\end{tabular}
	\caption{Underlying Probability Law (to be continued)}
	\label{tab1}
\end{table}

\begin{table}[htbp]
	\centering
		\begin{tabular}{lllllll}
		\hline \hline
			values & ... & $x_6$ & $x_7$ & $x_8$ & $x_9$ & $x_{10}$\\ 
		\hline \hline
			& ...  & 49.365.000 & 58.335.000 & 67.305.000 & 76.275.000 & 85.245.000\\
			$\mathbb{P}(X=x_i)$ & ... &  0.1 & 0.2 & 0.2 & 0.1 & 0.1\\
		\hline \hline
		\end{tabular}
	\caption{Continuation of Table \ref{tab1}}
	\label{tab1b}
\end{table}

\noindent Table \ref{tab1} shows the good performance of estimation the Zenga's discrete for size samples from $n=100$ to $n=1500$. Such sizes are comparable with those of sample survey from population counted in dozen of millions.  

\begin{table}[htbp]
	\centering
		\begin{tabular}{lllllll}
			\hline \hline
			Size & 100 & 200 & 500 & 750 & 1000 & 750\\
			\hline \hline
			ERM & $3.6 \ 10^{-3}$ & $-5.36 \ 10^{-3}$ & $10^{-3}$ & $-8.41 \ 10^{-4}$ & $4.56 \ 10^{-5}$ & $-1.44 10^{-3}$\\ 
			MSE & $6.4 \ 10^{-2}$ & $3.35 \ 10^{-2}$ & $2.49 \ 10^{-2}$ & $2.16 \ 10^{-2}$ & $1.9 \ 10^{-2}$ & $1.64 \ 10^{-2}$\\
			\hline \hline
		\end{tabular}
	\caption{Mean errors (ERM), Mean Square Errors (MSE)}
	\label{tab2}
\end{table}
 
\noindent Figure \ref{fig1} shows the pretty good asymptotic normality approximation of the centered and normalized empirical Zenga's estimator.

\begin{figure}
	\centering
		\includegraphics[width=1.00\textwidth]{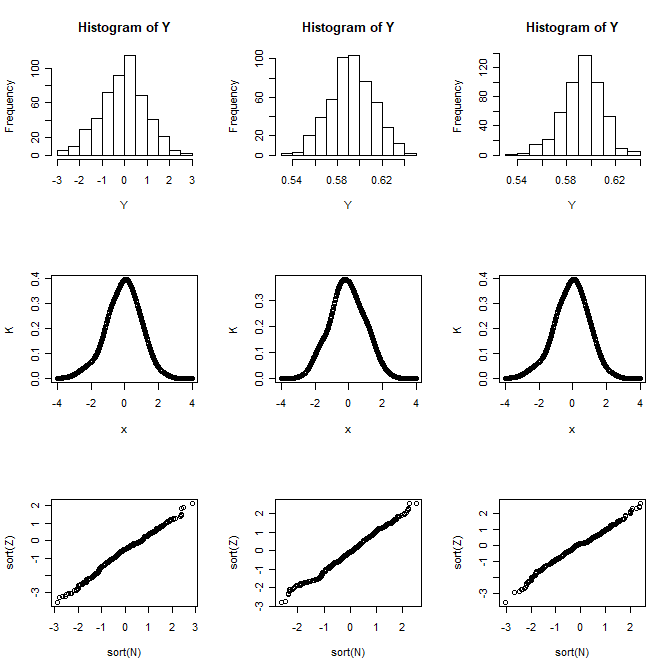}
	\caption{Histograms, Parzen Estimators and QQ-plots \newline for sample sizes $500$, $1000$ and $1500$ from left to right}
	\label{fig1}
\end{figure}

\bigskip \noindent \textbf{(B) Data-driven Applications}. \\

\noindent We use the income Data in Senegal (2001-2002) from the database related to \cite{esam2} . The incomes are given by households. We should use an adult-equivalence scale to consider to be able to compare households. The notion of adult-equivalence has already been described in \cite{loajas2016} and implemented on different sets of data, among them the data just described above. The data are available for the whole country (Senegal) and for the 10 areas given in the following order :\\

\noindent \textbf{(OA)} : Dakar, Diourbel, Fatik, Kaolack, Louga, Saint-Louis, Tamba, Thies, Ziguinchor, Kolda.\\

\noindent Dakar in the most urbanized area of Senegal and includes the capital of the country, named also after Dakar. It concentrated almost 23.1\% of the population.\\

\noindent The Zenga and the Gini index have been computed for the 11 areas from the aggregate data, and are display in Table \ref{tab2a} (continued in Table \ref{tab2b}).\\

\begin{table}
	\centering
		\begin{tabular}{llllllll}
		\hline \hline
			Index & Senegal & Dakar	& Diourbel	& Fatick 	& Kaolack & Louga 		& ..\\
		\hline \hline
			Zenga & 80.65		& 93.33	& 81.34			& 92.54		& 81.11		& 84.00			& ..\\
			Gini   &75.00     & 80.90 & 75.26     & 80.39   & 75.16		& 16.25		& ...\\
			\hline \hline
		\end{tabular}
	\caption{Zenga and Gini index measures for Senegal's administrative areas (2000), to be continued}
	\label{tab2a}
\end{table}

\begin{table}
	\centering
		\begin{tabular}{lllllll}
		\hline \hline
			Index & ... & Saint-Louis	& Tamba & Thies	& Ziguinchor	& Kolda\\
			\hline \hline
			Zenga & ... & 87.69				& 86.64	& 82.61	& 82.11				& 80.24\\
			Gini  & ... & 78.83				& 77.26 & 75.72 & 75.52				& 47.86\\
			\hline \hline
		\end{tabular}
	\caption{Continuation of Table \ref{tab2a}}
	\label{tab2b}
\end{table}

\noindent Through the values in theses tables, the 11 areas are ordered from the least inequality index to the greatest as follows :\\

\noindent \textbf{Ordering by Zenga's index} : Kolda (1),  Senegal (2), Kaolack (3), Diourbel (4), Ziguinchor (5), Thies (6),  Louga (7), Tamba (8),  Saint-Louis (9),
Fatick (10),  Dakar (11).\\

\noindent \textbf{Ordering by Gini's index} : Louga (1),  Kolda (2),  Senegal (3), Kaolack (4), Diourbel (5), Ziguinchor (6),  Thies (7), Tamba (8), Saint-Louis (9), Fatick (10),
Dakar (11).\\

\noindent These orderings are illustrated in Figure \ref{fig1}.\\

\begin{figure}
	\centering
		\includegraphics{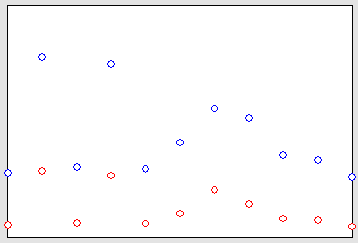}
	\caption{The areas are given in the horizontal line and are ordered according to the ranking \textbf{(AO)} above. Blue : Zenga's index. Regd : Gini's index}
	\label{fig1}
\end{figure}

\noindent The most striking fact is that the two index do not order the areas in an exact similar way. The most unfair areas (with the greatest values of the inequality index) are the same with the same ordering, form areas 8 to 11. From areas 1 to 7, the ordering is slightly changed but the case of Louga is remarkable. It is ranked first by Gini and seventh by Zenga.\\

\noindent One may think that the inequality should be greater in urban areas than in rural zone. Indeed we see that with the areas of Thies, Saint-Louis, Dakar. But Factik and Tamba are so urbanized areas. Investigating why the inequality indices (Both Zenga and Gini) are high should be investigated in accordance with local realities.\\

\noindent In this simple study, we are concerned with a large scale comparison studies between Zenga's and Gini's either but simulation studies or by theoretical investigations. This would be certainly in coming papers.

\section{Conclusion and perspectives} \label{zengaDist04}

\end{document}